\documentclass[9pt, twocolumn, twoside]{optica}
\usepackage{subfigure}

\usepackage{comment}

\setboolean{shortarticle}{false}
\setboolean{minireview}{false}

\dates{}
\doi{}
\setboolean{displaycopyright}{false}

\title{Focusing on Bandwidth: Achromatic Metalens Limits}

\author[1]{Federico Presutti}
\author[2,*]{Francesco Monticone}

\affil[1]{School of Applied and Engineering Physics, Cornell University, Ithaca, New York 14853, USA}
\affil[2]{School of Electrical and Computer Engineering, Cornell University, Ithaca, New York 14853, USA}

\affil[*]{Corresponding author: francesco.monticone@cornell.edu}

\begin{abstract}
Metalenses have shown great promise in their ability to function as ultracompact optical systems for focusing and imaging.
Remarkably, several designs have been recently demonstrated that operate over a large range of frequencies with minimized chromatic aberrations, potentially paving the way for ultrathin achromatic optics.
Here, we derive fundamental bandwidth limits that apply to broadband optical metalenses, regardless of their implementation.
Specifically, we discuss how the product between achievable time delay and bandwidth is limited in any time-invariant system, and we apply well-established bounds on this product to a general focusing system.
We then show that all metalenses designed thus far obey the appropriate bandwidth limit.
The derived physical bounds provide a useful metric to compare and assess the performance of different devices, and offer fundamental insight into how to design better broadband metalenses.
\end{abstract}

\begin{document}

\maketitle

\section{Introduction}

The field of metasurfaces holds the promise of a revolution in many areas of optics and photonics.
In principle, any optical system may be made flat and compact by replacing the conventional optics with ultrathin devices, with great potential benefits in terms of size, cost, and ease of fabrication \cite{ReviewKhorasaninejadCapasso, ReviewTsengTsai, ReviewLalanneChavel}.
While metasurfaces can achieve arbitrary wavefront transformations and may even be used for optical wave-based computing \cite{MetamaterialMath, MetasurfaceComputing}, ``metalenses'' specifically designed for focusing and imaging represent one of the most important classes of metasurfaces for practical applications.
One of the main challenges in this context is the realization of thin metalenses operating over a broad wavelength range, with minimized chromatic aberrations.
In conventional optics, it is possible to stack various lenses to correct chromatic aberrations \cite{BornWolfBook}, but at the price of making the overall system more bulky and costly.
It is therefore remarkable that, in recent years, different groups have demonstrated metalenses with a fixed focal length over a large range of wavelengths, some even surpassing the achromatic performance of conventional lens systems (at least for normal incidence) \cite{ChenCapasso, ChenCapasso2, KhorasaninejadCapasso,%
 ShresthaYu, YeRayYi, LinTsai, ChungMiller,%
 MohammadMenon, BanerjiMenon, ZhangCapasso, WangTsai, BalliHastings,%
 LinJohnson, ChengZhuang, YuZhang%
}.

Regardless of its implementation, a metalens achieves focusing by changing the phase of an incoming plane wave, with a phase profile that must vary radially according to the following equation \cite{ReviewKhorasaninejadCapasso}:
\begin{equation}
\varphi(r,\omega) = - \frac{\omega}{c} \left( \sqrt{F^2 + r^2} - F \right) ,
\label{eq:huygensphase}
\end{equation}
where $\omega$, $F$, $r$ and $c$ are the angular frequency, focal length, radial coordinate and speed of light, respectively.
In the general case, a spatial- and frequency-dependent phase profile, as in \eqref{eq:huygensphase}, can be Taylor expanded around a central frequency $\omega_c$:
\begin{align}
\varphi(r,\omega) = ~& \varphi(r, \omega_c) +
(\omega - \omega_c) {\left. {\frac{\partial \varphi(r, \omega)}{\partial \omega}} \right|_{\omega = {\omega _c}}}
\notag \\
&~ + \frac{1}{2} (\omega - \omega_c)^2 {\left. {\frac{\partial^2 \varphi(r, \omega)}{\partial \omega^2}} \right|_{\omega = {\omega _c}}} + \ldots
\label{eq:phaseexpansion}
\end{align}
where the latter two terms are the group delay and the group-delay dispersion.
As discussed by Chen et al.\ \cite{ChenCapasso}, when the focal length $F$ is frequency independent, \eqref{eq:phaseexpansion} contains no higher-order terms than the linear one.
Thus, to realize a perfectly achromatic lens, the design should implement (i) a suitable frequency-independent phase pattern $\varphi(r,\omega_c)$, (ii) a spatial pattern of group delay, and (iii) zero group-delay dispersion and higher-order terms.

Early examples of optical metalenses were based on deeply subwavelength ``meta-atoms'' (e.g.\ plasmonic dipole nano-antennas) operating near a scattering resonance to achieve the phase delay $\varphi(r,\omega_c)$ over the smallest possible thickness (a fraction of a wavelength), without any considerations of the linear and higher-order terms in \eqref{eq:phaseexpansion}.
While this strategy allows realizing arbitrarily thin metasurfaces, due to their unavoidable resonant nature, these devices were highly dispersive with large chromatic aberrations, and could only operate at either a single frequency or a discrete set of frequencies \cite{ReviewKhorasaninejadCapasso, ReviewTsengTsai, ReviewLalanneChavel}.
In contrast, the most recent designs at the time of writing have employed relatively thicker meta-structures (still on the order of a wavelength) that function essentially as microscopic waveguide segments.
The waveguiding approach does not depend on the phase delay obtained through near-resonant light interaction with a scatterer, but rather on the \emph{true time delay} obtained via guided-wave propagation, thus allowing significantly larger bandwidths.
Conceptually, this new approach is more similar to early examples of flat lenses at microwave frequencies, e.g., \cite{Pozar}, than to the first versions of modern optical metasurfaces based on resonant meta-atoms.

While the results obtained in recent works on broadband metalenses are remarkable, here we argue that there exists a strict physical bound on the chromatic properties of a metalens, which stems from the fact that it is not possible to impart an arbitrary group delay to a signal independently of its bandwidth.
Indeed, the delay-bandwidth product is limited in any linear, time-invariant system, as recognized in several works, \cite{MillerSlowLight, TuckerSlowLight, SlowLightBook}, and is strictly related to the thickness of the device.
Based on this concept, in the following we derive fundamental limits on the bandwidth of achromatic metalenses, and assess the performance of various existing designs against these limits.

\section{Results}

\emph{Time-bandwidth products} ---
Some attempts at identifying limits on the bandwidth of specific metalens designs have been recently made.
For example, Shrestha et al.\ \cite{ShresthaYu} have derived a bound on metalens bandwidths based on the range of dispersion properties covered by a meta-structure library, and Fathnan and Powell \cite{FathnanPowell} have derived bandwidth limits on low-frequency metasurfaces composed of printed-circuit impedance sheets.
Here, instead, we are interested in a fundamental limit, applicable in general to any metalens, regardless of its specific implementation.
With this goal in mind, we turn to the concept of delay-bandwidth, or time-bandwidth, product (TBP): the TBP of a device (a cavity, waveguide, etc.) is the product of the time delay, or interaction time, $\Delta T$, experienced by the signal, and the signal bandwidth $\Delta \omega$.
Wave physics imposes an upper bound on this quantity, which can be generally written as
\begin{equation}
\Delta T ~ \Delta \omega \leq \kappa ,
\label{eq:timebandwidthproduct}
\end{equation}
where $\kappa$ is a dimensionless quantity.
Bounds on the time-bandwidth product have been studied extensively in the field of slow light \cite{MillerSlowLight, TuckerSlowLight, SlowLightBook}.
In particular, as discussed below, different bounds have been derived in the literature under different assumptions, but $\kappa$ always depends on some general properties of the device, for example its length and refractive-index contrast of the materials involved.

\begin{figure}[tbp]
 \centering
 \includegraphics[width=0.95\columnwidth]{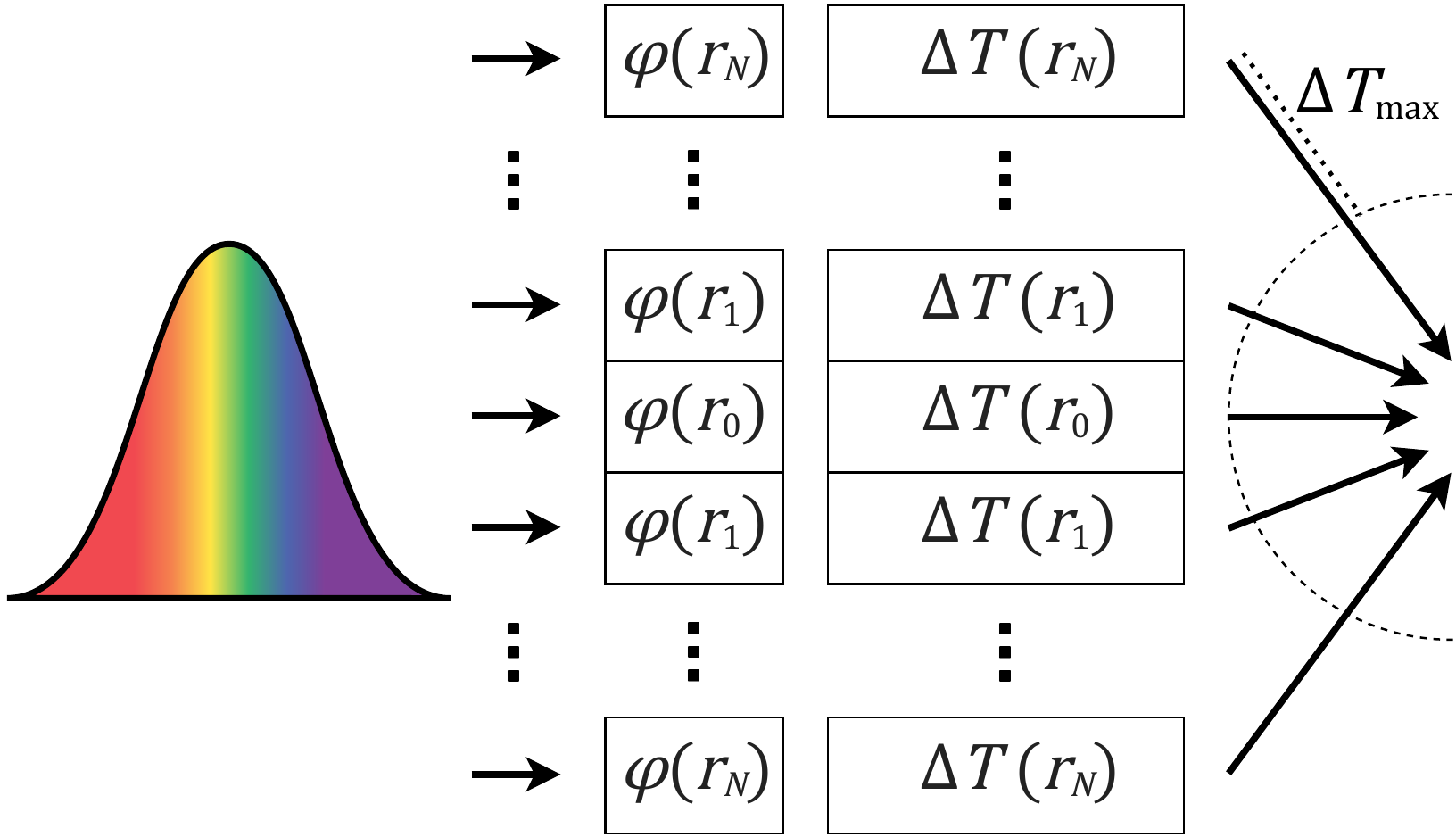}
 \caption{
  Delay-line model of a thin broadband metalens.
  Radially arranged delay lines provide a broadband signal the necessary group delay $\Delta T (r)$ to compensate for the difference in arrival times at the focus, while the phase pattern $\varphi (r)$ creates a spherical wavefront according to Eqs.~(\ref{eq:huygensphase}) and (\ref{eq:phaseexpansion}).}
 \label{fig:delaymodel}
\end{figure}

In order to apply the concept of TBP and the associated bounds to our problem, we treat a metasurface lens as composed of one-dimensional slow-light devices or delay lines.
More specifically, we consider a rotationally symmetric radial array, or continuum, of delay-line buffers, such that the incident wave is delayed as a function of radius, as illustrated in Fig.~\ref{fig:delaymodel}.
According to \eqref{eq:phaseexpansion}, each delay-line buffer must impart a suitable group delay, dispersionless over the given band, to compensate for the difference in the arrival times of wavepackets at the focus.
The lens must also impart a frequency-independent phase pattern, $\varphi(r, \omega_c)$, to create a spherical wavefront at the output.
This phase pattern can be implemented independently of the group-delay requirement, using, for example, the concept of geometric phase, as shown in Ref.~\cite{ChenCapasso}.
We would like to note that the 1D model in Fig.~\ref{fig:delaymodel} is an approximation since a metasurface with finite thickness is not strictly a one-dimensional device; however, we expect that light normally incident on metasurfaces of thicknesses on the order of one wavelength would acquire time delay predominantly through longitudinal propagation, with little propagation along the lateral (radial) direction.
For this reason, in what follows, we only consider metasurfaces with thickness smaller than about five free-space wavelengths.
We further discuss this approximation and its implications in Section \ref{sec:conclusion}.

Consider a lens with radius $R$ and focal distance $F$.
From \eqref{eq:huygensphase} and \eqref{eq:phaseexpansion}, the required relative group delay imposed by the lens at a radial position $r \leq R$ is
\begin{align*}
\Delta T(r) &=
\frac{\partial \varphi}{\partial \omega} (r) - \frac{\partial \varphi}{\partial \omega} (R) \\
&= \frac{1}{c} \left( \sqrt{F^2 + R^2} - \sqrt{F^2 + r^2} \right) .
\end{align*}
(In most cases, the group delay is equal to the actual time delay experienced by the signal, except in the presence of anomalous dispersion near resonances, which is, however, not a case of interest in this context, due to the strong absorption that unavoidably accompanies anomalous dispersion in passive systems \cite{JacksonBook}.)
The greatest delay is required at the center ($r=0$) to compensate for the additional time taken by a signal arriving from the edge ($r=R$), such that the lens design must achieve
\begin{align}
\Delta T_\text{max} = \frac{F}{c} \left( \sqrt{1 + (R/F)^2} - 1 \right),
\label{eq:Tmax}
\end{align}
as illustrated in Fig.~\ref{fig:delaymodel}.
This defines our required time delay in the time-bandwidth product in \eqref{eq:timebandwidthproduct}.

Then, using the numerical aperture definition
\begin{align}
\text{NA} = n_b \sin \theta = n_b \sin \left[\arctan \left( \frac{R}{F} \right) \right] ,
\label{eq:NA}
\end{align}
where $n_b$ is the background refractive index, and the identity $1 + \tan^2 (\arcsin x) = 1 / (1 - x^2)$,
we can use (\ref{eq:timebandwidthproduct}) and (\ref{eq:Tmax}) to set a limit on the lens' bandwidth based on its numerical aperture and geometrical properties:
\begin{align}
\Delta \omega \leq \frac{\kappa c}{F \left( \sqrt{1 + (R/F)^2} - 1 \right)}
= \frac{\kappa c \sqrt{1 - (\text{NA}/n_b)^2}}{F \left(1 - \sqrt{1 - (\text{NA}/n_b)^2} \right)} .
\label{eq:bandwidthLimit}
\end{align}

As mentioned above, different values of the upper bound $\kappa$ have been derived for different general classes of devices.
Thus, depending on the type of metasurface (whether it is based on resonant meta-atoms or waveguiding structures), we can apply the appropriate TBP bound and derive the relevant bandwidth limit.
We now identify three relevant cases, covering all types of metasurfaces.

(i) If the metasurface has deeply subwavelength thickness, the only way to impart the necessary phase/time delay to incoming light is by interaction with resonant scattering meta-atoms.
In this case, coupled-mode theory provides a geometry- and material-independent time-bandwidth product for a single-mode resonator (e.g.,\ see \cite{MannAlu}):
\begin{equation}
\kappa = 2.
\label{eq:tbpSR}
\end{equation}
In this case, \eqref{eq:timebandwidthproduct} becomes an equality and it may also be written in terms of the \textit{Q} factor of the resonant meta-atoms, $\Delta T = 2 Q /\omega_c$, as recognized in Refs.~\cite{ReviewLiangKrauss, DiffractiveOptics}.

(ii) For thick metasurfaces based on inclusions acting as waveguides, the previous limit clearly does not apply, since wave-guiding structures cannot be treated as individual resonators.
Tucker et al.\ \cite{TuckerSlowLight} provide a generally applicable time-bandwidth limit valid for any one-dimensional, lossless, dielectric device that may be treated as a waveguide (the limit is strictly valid only if the fractional bandwidth is not too large, i.e., smaller than unity).
The value of the upper bound $\kappa$ is given by
\begin{equation}
\kappa = 2 \pi \frac{L}{\lambda_c} (n_\text{max} - n_\text{min}),
\label{eq:tbpTucker}
\end{equation}
where $n_\text{max}$ and $n_\text{min}$ are the average and minimum \emph{effective} refractive indices of the device.
This effective index $n(\omega)$ is defined by an effective dispersion relation for the mode of interest in the structure, $\beta(\omega) = \omega n(\omega) / c$, as if the structure was homogeneous.
$n(\omega)$, $n_\text{max}$ and $n_\text{min}$ are generally different from the material indices that make up the device.
However, in the case of a one-dimensional dielectric waveguide segment (or coupled segments, as in \cite{ChenCapasso}), which is the case of interest for most modern metasurfaces, we can take $n_\text{max}$ and $n_\text{min}$ as the refractive index of the dielectric material composing the waveguide and of the surrounding medium, respectively, because the guided-mode dispersion converges to the light line of the low-index material at low frequency, and to the light line of the high-index material at high frequency.
We thus replace the difference term in \eqref{eq:tbpTucker} with $\Delta n = n_\text{max} - n_b$.

(iii) Finally, Miller \cite{MillerSlowLight} provides a similar, but much more general, time-bandwidth limit valid for a very broad class of one-dimensional structures (not necessarily dielectric) acting as delay lines:
\begin{equation}
\kappa = \frac{\pi}{\sqrt{3}} \frac{L}{\lambda_c} \eta_{\max} ,
\label{eq:tbpMiller}
\end{equation}
where $\eta_{\max} = |(\varepsilon_{\max} - \varepsilon_b) / \varepsilon_b|$ is the device's maximum contrast in relative permittivity, with respect to the surrounding medium's permittivity $\varepsilon_b$, at any frequency within the band of interest and at any position within the structure.
The limit is very general, as it is independent of the device design, and is not based on the simplifying assumptions used in Tucker's limit (lossless dielectric materials and well-defined group velocity).
The limit strictly applies if the device length $L \gg \lambda_c$, where $\lambda_c$ is the band-center wavelength in the background medium, and if the fractional bandwidth is not too large.
However, in practice, we have verified that, if $L$ is just a few times larger than the longest wavelength in the device, Miller's limit seems to apply, that is, it is consistent with Tucker's limit (which has no assumptions on length).
If they both apply, Miller's limit is close to Tucker's, and exceeds it if $\varepsilon_\text{max} \gtrapprox 6$.

These bounds on the time-bandwidth product, within their limits of applicability, may be combined with \eqref{eq:bandwidthLimit} to obtain an upper bound on the bandwidth of achromatic metalenses.
Essentially all types of metalenses, for any thickness (smaller than a few wavelengths) and material composition, are covered by the three TBP bounds outlined above, leading to the following bandwidth limits:

(i) for ultra-thin metasurfaces based on resonant meta-atoms [from (\ref{eq:tbpSR})],
\begin{equation}
\Delta \omega \leq \frac{2 c}{F} \Theta \! \left( \frac{\text{NA}}{n_b} \right) ,
\label{eq:limitSR}
\end{equation}

(ii) for waveguide-based dielectric metasurfaces [from (\ref{eq:tbpTucker})],
\begin{equation}
\Delta \omega \leq \omega_c \frac{L \Delta n}{F} \Theta \! \left( \frac{\text{NA}}{n_b} \right) ,
\label{eq:modifiedTucker}
\end{equation}

(iii) for generic metasurfaces (not necessarily dielectric and lossless) of thickness larger than the wavelength [from (\ref{eq:tbpMiller})],
\begin{equation}
\Delta \omega \leq \frac{\omega_c}{2 \sqrt{3}} \frac{L \eta_\text{max}}{F} \Theta \! \left( \frac{\text{NA}}{n_b} \right) ,
\label{eq:limitMiller}
\end{equation}
where we replaced $\omega_c = 2 \pi c / \lambda_c$ and
\[
\Theta \! \left( \frac{\text{NA}}{n_b} \right) = \frac{\sqrt{1-(\text{NA}/n_b)^2}}{1-\sqrt{1-(\text{NA}/n_b)^2}} .
\]

\begin{figure}[tbp]
 \begin{flushleft} \large{\textbf{a)} \vspace{-\baselineskip}} \end{flushleft}
 \subfigure {
 \includegraphics[width=0.95\columnwidth]{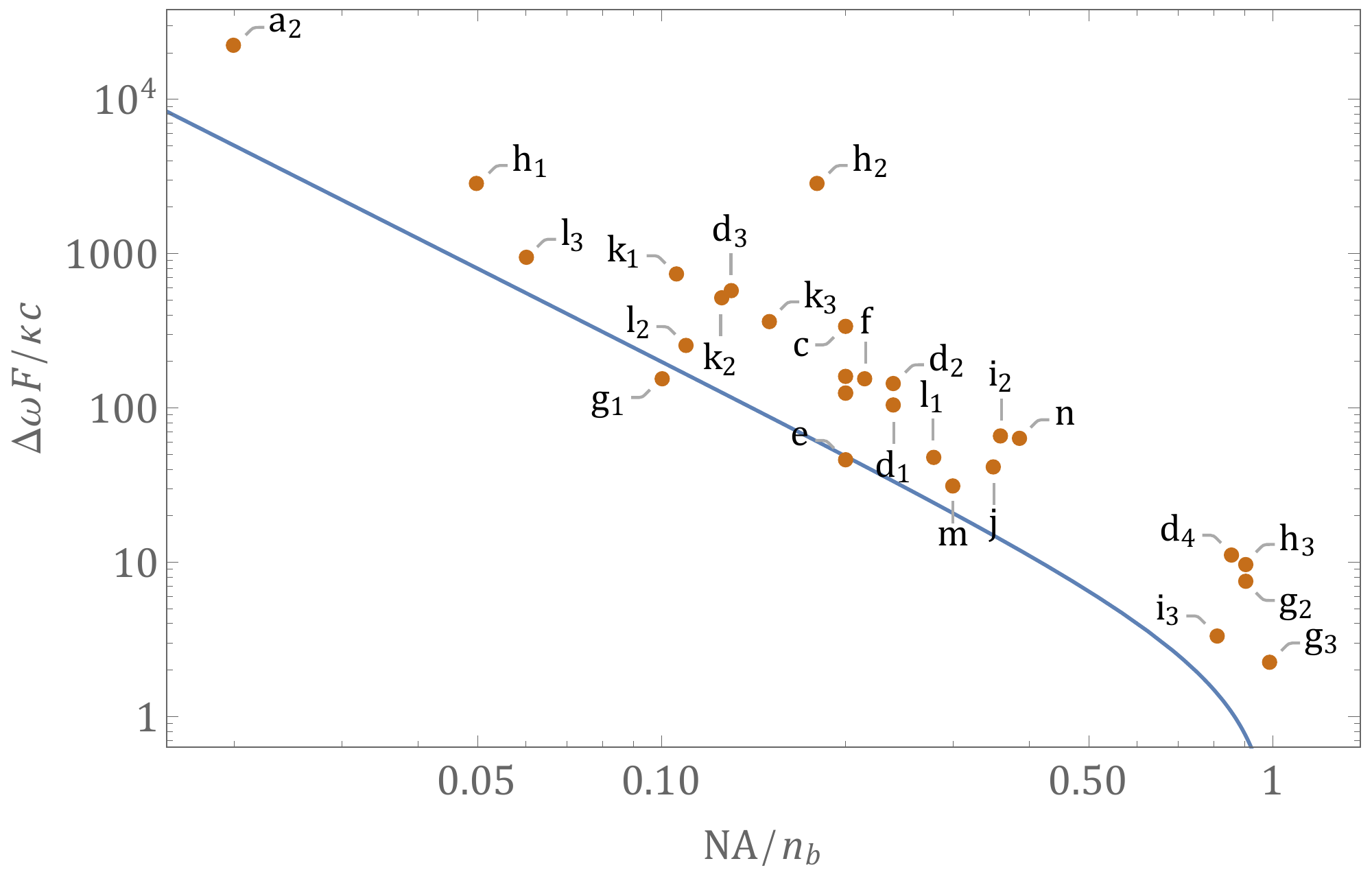}
 \label{fig:LimitPlotsA}
 }
 \begin{flushleft} \large{\textbf{b)} \vspace{-\baselineskip}} \end{flushleft}
 \subfigure{
 \includegraphics[width=0.95\columnwidth]{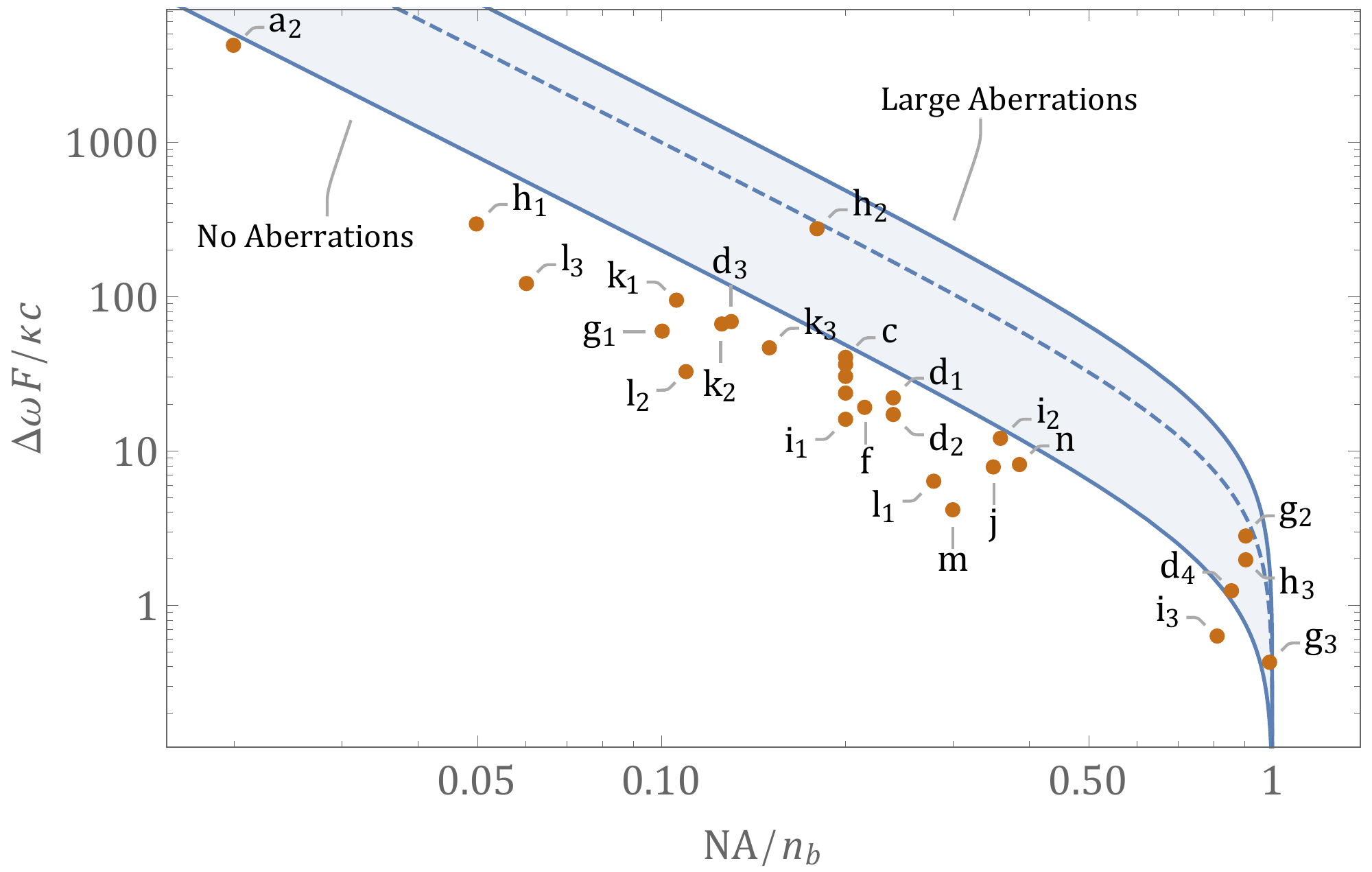}
 \label{fig:LimitPlotsB}
 }
 \caption{
  Comparisons of published achromatic metalens designs against the derived bandwidth limits.
  (a) Limit based on the single-resonator TBP bound, given by \eqref{eq:limitSR}.
  Not surprisingly, most data points exceed this bound except for some thin devices.
  (b) Limit based on Tucker's TBP bound, given by \eqref{eq:modifiedTucker}.
  Each data point in the plots represent a single design, with each label corresponding to a specific row of Table ~\ref{fig:LimitTable}.
  The performance of each metalens is represented in terms of numerical aperture and bandwidth.
  In order to compare vastly different designs against the bandwidth bounds, the bandwidths are normalized by $c$, $F$ and the corresponding $\kappa$ (see \eqref{eq:bandwidthLimit}).
  In both panels, the lowest blue curve represents the function $\Theta(\text{NA} / n_b) = c / F \Delta T$, where $\Delta T$ is the required time delay for ideal operation, given by \eqref{eq:Tmax}.
  Fig.~\ref{fig:LimitPlotsB} includes both the upper bound for ideal metalenses with no aberrations, $\Delta T_{\text{err}} = 0$ ($\Theta$, lower solid blue curve), and bounds for highly aberrated metalenses with an error $\Delta T_{\text{err}} = 0.8 \Delta T$ ($\Theta/0.2$, dashed blue curve) and $\Delta T_{\text{err}} = 0.9 \Delta T$ ($\Theta/0.1$, upper solid blue curve), which corresponds to low values of Strehl ratio according to \eqref{eq:Strehl}.
  All design parameters and bandwidth values are given in Table~\ref{fig:LimitTable}.
 }
 \label{fig:LimitPlots}
\end{figure}

We compared these bandwidth limits to various broadband metalens designs available in the literature.
Comparisons are shown in Fig.~\ref{fig:LimitPlots}, using suitably normalized quantities, and are tabulated in Table~\ref{fig:LimitTable}.
What immediately stands out is that the limits correctly predict the expected performance trend: for larger numerical aperture the achievable bandwidth shrinks, because the required maximum time delay, $\Delta T_\text{max}$ in \eqref{eq:Tmax}, rapidly increases (and diverges at $\text{NA}/n_b = 1$).

Not surprisingly, as seen in Fig.~\ref{fig:LimitPlotsA}, only a few thin, subwavelength, metalenses obey the bandwidth limit based on the single-resonator TBP given in \eqref{eq:tbpSR}.
Instead, all the metalens designs obey our limits based on Tucker's or Miller's TBP (see Fig.~\ref{fig:LimitPlotsB} and Table 1), including recent ultrabroadband metalenses obtained using free-form all-area optimization and inverse design \cite{ChungMiller}.

At this point, it is important to note that, although all the designs considered claim achromatic performance, some have non-negligible focal length variations, or do not disclose the exact field profile at the focal plane and the associated level of aberrations.
The focal field profile is relevant because the bandwidth limit derivation above assumes a diffraction-limited lens with no aberrations, i.e., Strehl ratio $S=1$.
The Strehl ratio is a measure of the wavefront aberration, defined as the ratio of the peak focal spot intensity to the maximum attainable intensity of an ideal lens.
A metalens that does not achieve diffraction-limited focusing with $S=1$ across the nominal operational bandwidth is not implementing the phase/time delay assumed above and, therefore, may surpass the bounds since the requirements are somewhat relaxed.

To quantify the effect of aberrations on our bandwidth bounds, we assume that the aberrations are not too large, which is the scenario of interest for imaging applications.
In this case, the Strehl ratio is approximately independent of the nature of the aberration and, according to the ``extended Mar\'echal approximation,'' it can be estimated from the variance of the wavefront deformation with respect to an ideal spherical wavefront:
\[
S \approx \mathrm{e}^{- \left( k_0 \sigma \right)^2},
\]
where $\sigma$ is the standard deviation of the spatial wavefront deformation and $k_0$ is the free-space wavenumber \cite{BornWolfBook, WavefrontAberrationTheory, AdaptiveOptics}.
In addition, the maximum peak-to-peak deformation of the wavefront can be related to the standard deviation as $\Delta W_\text{max} \approx \alpha \sigma$, where the factor $\alpha$ depends on the the type of aberration, and for a mixture of low-order aberrations (defocusing, etc.) $\alpha \approx 4.5$ \cite{WavefrontAberrationTheory}.
This spatial error corresponds to a maximum phase error $\Delta \varphi_\text{max} = \Delta W_\text{max} k_0$.
Thus, if a less-than-unity Strehl ratio is tolerated within the operational bandwidth of the metalens, an error in the implemented phase profile would be acceptable, which in turn would relax the requirements on the time delay $\Delta T$ and the associated bandwidth bounds.
In particular, assuming no phase errors at the central frequency, if the implemented time delay is incorrect by a maximum amount $\Delta T_{\text{err}}$, \eqref{eq:phaseexpansion} indicates that the phase profile would be incorrect by an amount $\Delta \varphi_\text{max} =(\omega - \omega_c) \Delta T_{\text{err}}$ at a certain frequency $\omega$, corresponding to a Strehl ratio:
\begin{equation}
S \approx \exp \! \left( - \! \left(\alpha^{-1}(\omega - \omega_c) \Delta T_{\text{err}} \right)^2 \right).
\label{eq:Strehl}
\end{equation}
Based on these considerations, it is then possible to approximately account for aberrations in our bandwidth bounds by substituting $\Delta T$ with $\Delta T - \Delta T_{\text{err}}$ in \eqref{eq:timebandwidthproduct} and expressing $\Delta T_{\text{err}}$ in terms of $S$ by inverting \eqref{eq:Strehl}.
This leads to a looser bandwidth bound if the Strehl ratio decreases, suggesting that a relaxation of the imaging performance of the metalens allows for a broader bandwidth, as expected.

Since we do not have access to the field profiles of all the metalenses considered in the literature, in Fig.~\ref{fig:LimitPlotsB} we include both the bound for ideal metalenses with no aberrations (lower solid blue curve), and a bound for highly aberrated metalenses with an error $\Delta T_{\text{err}} = 0.9 \Delta T$ (upper solid blue curve), corresponding to low values of Strehl ratio according to \eqref{eq:Strehl}.
Most published metalens designs are below the bound for ideal metalenses, with only a handful of designs exceeding this limit.
However, the latter are all bound by the limit for aberrated metalenses with $\Delta T_{\text{err}} = 0.8 \Delta T$ (dashed blue curve), corresponding to a typical Strehl ratio $< 0.5$ away from the central wavelength, which is consistent with the published results (we note that since the nominal $\Delta T$ depends on $F$ and NA, according to \eqref{eq:Tmax}, the resulting Strehl ratio also depends on these quantities).
Thus, in principle, even broader bandwidths could be achievable, but only at the expenses of even higher aberrations and lower focal spot intensity.

\begin{figure}[tbp]
 \includegraphics[width=1.03\columnwidth]{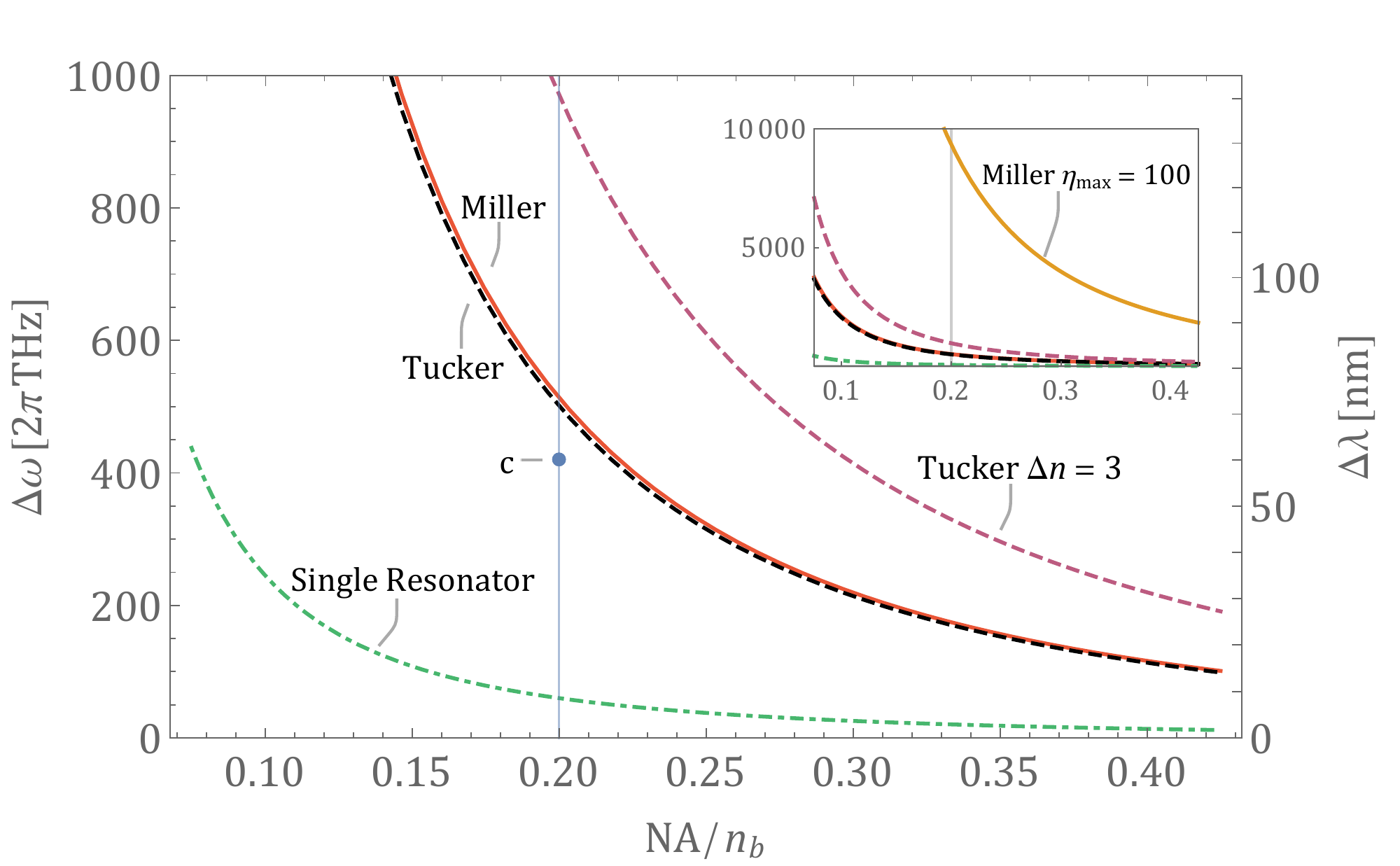}
 \caption{
  Different bandwidth limits compared to the performance of a specific metalens design, from Ref.~\cite{KhorasaninejadCapasso} (central wavelength $\lambda_c= 518$ nm):
  Single-resonator limit (\eqref{eq:tbpSR}, dot-dashed green curve);
  Tucker's limit (\eqref{eq:tbpTucker}, dashed black) and Miller's limit (\eqref{eq:tbpMiller}, solid red) using the actual refractive-index/permittivity contrast considered in Ref.~\cite{KhorasaninejadCapasso}; design-independent Tucker's limit (dashed purple) with the highest refractive index for lossless dielectrics available at optical frequency, $n \approx 4$.
  The inset includes the same curves and an additional curve (solid orange) for a design-independent version of Miller's limit with the highest permittivity contrast available at optical frequency $\eta \approx 100$ \cite{MillerSlowLight} (which may include metallic materials, loss and gain).
 }
 \label{fig:LimitPlots_example}
\end{figure}

Finally, in Fig.~\ref{fig:LimitPlots_example}, we show an example of how a specific metalens design (from Ref.~\cite{KhorasaninejadCapasso}) compares with the bandwidth limits described above, considering the case of no aberrations for simplicity.
This metasurface, which is based on dielectric waveguide segments, has a much larger bandwidth than what would be achievable using a single-resonator-based design, as expected.
Instead, its bandwidth performance is not too far from the appropriate upper bound (either \eqref{eq:tbpTucker} or \eqref{eq:tbpMiller}) based on the employed materials and thickness.
In other words, the dielectric metalens is using its thickness and refractive-index contrast almost optimally.
Fig.~\ref{fig:LimitPlots_example} also shows a design-independent version of both Tucker’s and Miller's limit using the highest refractive-index and permittivity contrast available at optical frequencies, for lossless dielectrics and generic materials, respectively.
Further details are discussed in Section \ref{sec:conclusion}.

\emph{Bandwidth limits on reflection suppression} ---
For the sake of completeness, we briefly discuss another important trade-off, between the bandwidth of operation of a metalens and its transmission efficiency.
The ability to transmit energy efficiently requires, at a minimum, that the reflections are minimized, namely, that the metalens is impedance matched with respect to the medium in which the incident wave propagates (usually air or a transparent substrate).
While it is always possible to design a lossless anti-reflection coating to achieve identically zero reflection (ideal impedance matching) at a single frequency, a fundamental trade-off exists between the reflection reduction and the continuous bandwidth over which this reduction can be achieved.
This fundamental limit on broadband impedance matching is known as the Bode-Fano limit \cite{BodeFano}, which has been used for decades in microwave engineering, but it applies equally well at optical frequencies \cite{MonticoneAlu}.
This bound depends uniquely on the linearity, passivity, time-invariance, and causality of the scattering system, and, most importantly, is independent of the employed anti-reflection coating, regardless of its complexity (the matching structure is only assumed to be lossless).

In order to apply the Bode-Fano limit to the problem under consideration, we approximate the metalens as a thin homogeneous slab with refractive index equal to the average refractive index $n_\text{avg}$ of the materials composing the structure.
This is clearly a coarse approximation, but it allows us to get some general insight on this design trade-off.
In addition, we assume that we operate in the most favorable condition for impedance matching, that is, we assume that the central frequency corresponds to a Fabry-Perot resonance of the slab, at which the reflection coefficient automatically goes to zero.
The slab thickness $L$ is assumed to be smaller or comparable to the wavelength.
Under these approximations, the limit is given by \cite{MonticoneAlu}
\begin{equation}
\frac{\Delta \omega}{\omega_c} \leq \frac{1}{L/\lambda_c (\varepsilon - \varepsilon_b)} \left[ \log \frac{1}{| \Gamma |} \right]^{-1} ,
\label{eq:bodefano}
\end{equation}
where $\varepsilon=n_\text{avg}^2$ and $\Gamma$ is the in-band reflection coefficient.
If the equal sign is used, \eqref{eq:bodefano} represents the optimal trade-off between bandwidth and reflection reduction.

\begin{figure}[tbp]
 \includegraphics[width=0.95\columnwidth]{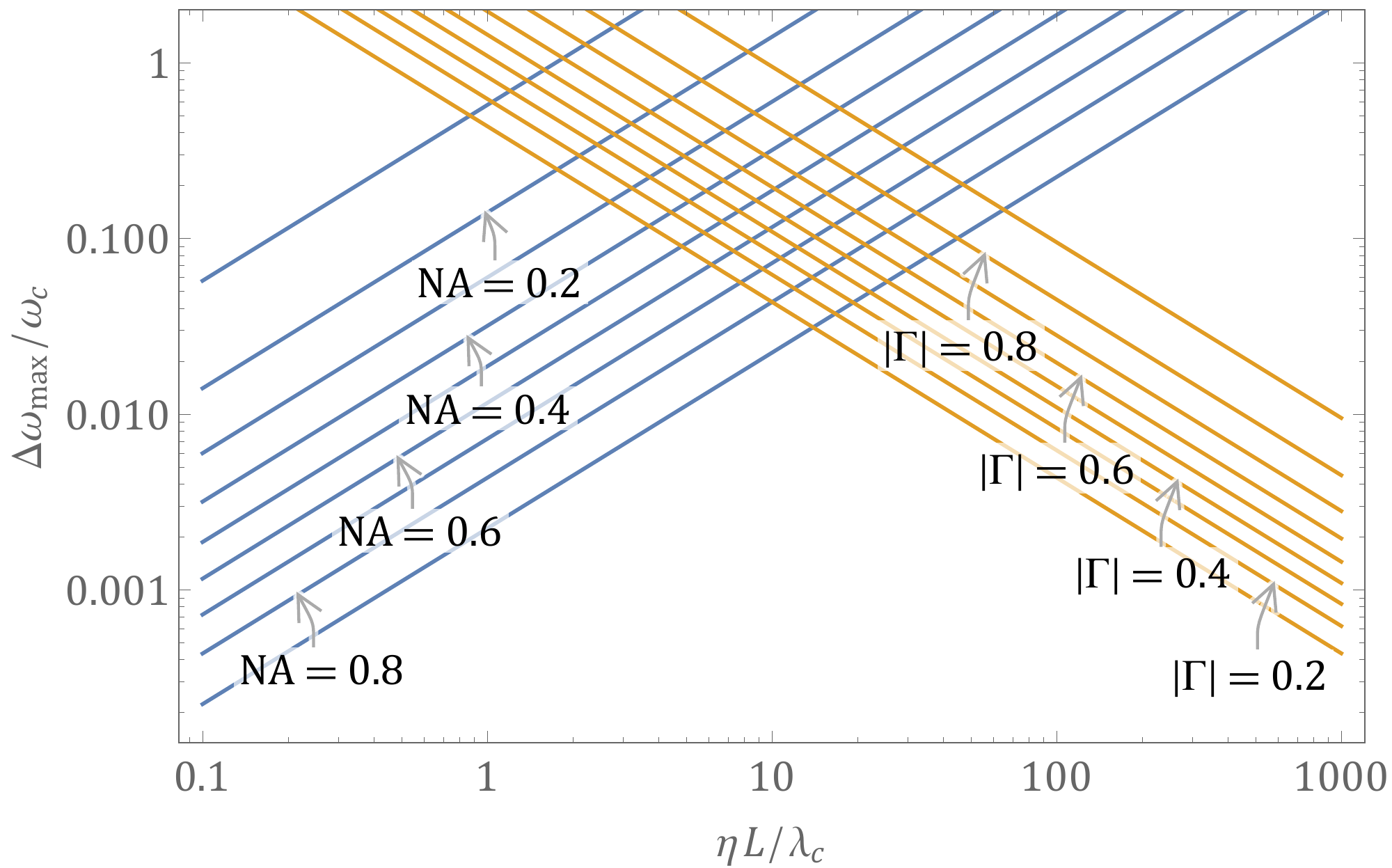}
 \caption{
  Comparison of the bandwidth limit for achromatic performance (blue), based on Miller's TBP, \eqref{eq:limitMiller}, and the Bode-Fano bandwidth limit on reflection reduction, \eqref{eq:bodefano} (orange), as a function of the product of permittivity contrast and normalized thickness: $\eta L/\lambda_c$.
  The limits are compared for various values of NA and in-band reflection coefficient $|\Gamma|$.
  As an example, we considered a metalens with $F = 100 \lambda_c$ and $\varepsilon_b = 1$.
  In order to apply the Bode-Fano limit to the considered problem, we assumed that the metasurface is homogeneous with $\eta = \varepsilon - \varepsilon_b = \eta_\text{max}$.
  For a given value of NA and $|\Gamma|$ there is an optimal value of $\eta L/\lambda_c$, where the two limits intersect, that maximizes $\Delta \omega$.
 }
 \label{fig:bandwidth_reflection_NA}
\end{figure}

Depending on the application under consideration, the maximum bandwidth over which a metalens can operate depends on both the achromatic focusing limit derived above and the impedance-matching limit.
Interestingly, it is immediately clear that \eqref{eq:bodefano} is inversely proportional to the thickness and refractive-index/permittivity contrast of the device, while \eqref{eq:modifiedTucker} and \eqref{eq:limitMiller} are directly proportional to these quantities.
This suggests the existence of a fundamental trade-off between the ability to reduce reflections with an anti-reflection coating (maximizing transmission efficiency) and the ability to minimize chromatic aberrations for a metasurface operating over a broad continuous bandwidth.
This trade-off is represented in Fig.~\ref{fig:bandwidth_reflection_NA}: thicker devices or larger refractive-index contrast lead to wider bandwidths for achromatic operation (blue curves), but narrower bandwidths over which the reflection coefficient can be reduced to a certain level (orange curves).
In other words, achieving achromatic performance over a wider band requires a larger $\eta L/\lambda_c$, which, however, increases the minimum reflectance achievable over that band, as expected.
If both efficiency and achromatic performance are equally important, an optimal value of $\eta L/\lambda_c$ may be identified depending on the specific application under consideration.

\section{Discussion and Conclusion}
\label{sec:conclusion}

Considering the bandwidth limits on achromatic metalenses discussed above, one may wonder what type of metalens design, for a fixed refractive-index/permittivity contrast and thickness, can get closest to the limit and why.

Interestingly, for a certain refractive-index contrast, a metasurface design based on suitable dielectric waveguide segments seems to directly provide a way to realize performance close to the upper bound for the given thickness.
Indeed, the guided-mode dispersion of a dielectric waveguide converges to the light line of the low-index material at low frequency, and to the light line of the high-index material at high frequency.
This provides an intermediate frequency window with low group velocity and locally linear dispersion that is automatically close to the optimal linear dispersion considered by Tucker et al.\ \cite{TuckerSlowLight} for an ideal delay line for that level of contrast.

It is therefore not surprising that, even when considering free-form all-area optimization of dielectric metalenses as in \cite{ChungMiller}, the optimization tends to create a spatial distribution of material with ``channels'' that resemble waveguide segments.
It is also not surprising that many of the designs we considered are relatively close to the limit, as shown in Fig.~\ref{fig:LimitPlotsB}, since many make use of the available length and refractive-index contrast almost optimally.
Considering the largest available refractive index for a transparent material at optical frequency, which is around three to four in silicon and germanium, would certainly provide a wider bandwidth, but not an order-of-magnitude improvement with respect to metasurfaces fabricated with lower values of refractive index.
Fig.~\ref{fig:LimitPlots_example} (purple dashed curve) shows the bandwidth limit for this maximum value of refractive index, $n=4$, compared to the bandwidth of the metalens in Ref.~\cite{KhorasaninejadCapasso}.
Such a bandwidth limit provides a design-independent upper bound for transparent dielectric metasurfaces, which only depends on the thickness and the desired focal length and NA.
Since it is unlikely that a much larger natural refractive-index contrast could be achieved at optical frequencies, the only way to improve the bandwidth performance using transparent materials is to consider longer devices, or overcome the limit by breaking its main assumptions, for example, time-invariance, a possibility that will be the subject of future works.

Considering much longer metalenses may also break the assumption of one-dimensionality on which the limits above are based (see Fig.~\ref{fig:delaymodel} and related discussion).
An example of this is the broadband metalens in Ref.~\cite{LinJohnson}, designed through free-form all-area optimization, whose thickness is more than five free-space wavelengths (and even longer considering the wavelength within the metasurface structure).
Thus, this metalens cannot be considered an array of one-dimensional delay lines as in Fig.~\ref{fig:delaymodel} since lateral propagation can no longer be neglected.
Indeed, this thick metasurface manages to surpass our bounds to some degree, with relatively small aberrations.
In general, we expect that thicker metasurfaces or stack of metasurfaces, with a thickness of several wavelengths, may be designed to optimally take advantage of the two- or three-dimensionality of the system, increasing the path a wavepacket travels laterally, not just longitudinally, which would in turn lead to wider achievable bandwidths.
In this context, we believe that all-area optimization is critical to fully take advantage of the whole available \emph{volume}.

Finally, one may also wonder whether it would be possible to artificially increase the maximum available refractive index by realizing an engineered metamaterial with effective index larger than the one of the constituent materials.
However, if the thickness of the metalens is limited to approximately a wavelength or few wavelengths, the meta-atoms must be very small to actually form an effective homogeneous metamaterial, and not act as a discrete arrangement of elements.
If we choose, for example, the size of the meta-atoms to be $d \approx \lambda/10$, a dielectric meta-atom would be largely off-resonance even considering the largest refractive index, $n \approx 4$ (the first resonance of a high-index dielectric sphere is of magnetic dipolar type, and it occurs when $d \approx \lambda/n $).
As a result, the effective permittivity would not be too different from the average between the permittivity of the inclusions and of the background, following standard mixing formulas for non-resonant meta-atoms (e.g.,\ see \cite{TretyakovModeling}).
Using plasmonic materials would allow realizing deeply subwavelength resonant meta-atoms and, therefore, a metamaterial with much larger effective permittivity.
This would, however, be accompanied by large Lorentzian dispersion around an unavoidable absorption peak, which would greatly reduce the bandwidth and efficiency of the device.
Still, Miller's limit in \eqref{eq:limitMiller}, which is based on the \emph{magnitude} of the permittivity contrast, does not preclude the possibility of achieving better bandwidth performance at optical frequencies by using metallic materials, for which the contrast can be as high as $\eta_{\max} \approx 100$ at near-infrared frequencies.
In theory, this would allow an order-of-magnitude improvement in bandwidth, as seen in the inset of Fig.~\ref{fig:LimitPlots_example} (orange curve), where we show Miller's limit for $\eta_{\max} \approx 100$.
This provides a ultimate upper bound on the bandwidth of optical metalenses that may include any available material.
However, there is no guarantee that this limit is tight, namely, that it could be achieved with a physical design \cite{MillerSlowLight}.

To conclude, we believe that the fundamental bandwidth limits presented in this Article will prove useful to the many research groups working on metasurfaces to assess and compare the performance of different devices, and may offer fundamental insight into how to design broadband achromatic metalenses for different applications.




\bibliographystyle{osajnl}
\bibliographystyle{jabbrv}
\bibliography{%
 ReviewKhorasaninejadCapasso,%
 ReviewTsengTsai,%
 ReviewLalanneChavel,%
 MillerSlowLight,%
 TuckerSlowLight,%
 SlowLightBook,%
 ChenCapasso,%
 ChenCapasso2,%
 KhorasaninejadCapasso,%
 ShresthaYu,%
 YeRayYi,%
 LinTsai,%
 ChungMiller,%
 MohammadMenon,%
 BanerjiMenon,%
 ZhangCapasso,%
 WangTsai,%
 BalliHastings,%
 LinJohnson,%
 ChengZhuang,%
 YuZhang,%
 BodeFano,%
 MonticoneAlu,%
 TiO2_1,%
 TiO2_2,%
 Palik,%
 GaSb,%
 MetamaterialMath,%
 MetasurfaceComputing,%
 Pozar,%
 FathnanPowell,%
 MannAlu,%
 TretyakovModeling,%
 ReviewLiangKrauss,%
 DiffractiveOptics,%
 WavefrontAberrationTheory,%
 BornWolfBook,%
 JacksonBook,%
 AdaptiveOptics%
}

\begin{table*}[htbp]
\centering
\resizebox{\textwidth}{!}{
 \setlength\arrayrulewidth{0.75pt}
 \begin{tabular}{|c|c|c|c|c|c|c|c|c|c|c|}
 \hline
 Design & NA (NA$_\text{eff}$) & $F$ / $\mu$m & $\lambda_{\min}$ / nm & $\lambda_{\max}$ / nm & $L$ / nm & $\varepsilon_{r\text{max}}$ &
 $\Delta \omega$ / THz & SR / THz & Miller / THz & Tucker / THz \\
 \hline
a$_1$ \cite{ChenCapasso} &
 0.20 & 63 & 470 & 670 & 600 & 6.5 & 1200 &
  462 &
  2500 (12\,500) &
  2440 (12\,200) \\
a$_2$ \cite{ChenCapasso} &
 0.02 & 11\,000 & 470 & 670 & 600 & 6.5 & 1200 &
  272 &
  1480 (7380) &
  1440 (7200) \\
b \cite{ChenCapasso2} &
 0.20 & 67 & 460 & 700 & 600 & 6.5 & 1400 &
  434 &
  2340 (11\,700) &
  2280 (11\,400) \\
c \cite{KhorasaninejadCapasso} &
 0.20 & 485 & 490 & 550 & 890 & 6.5 & 419 &
  60.0 &
  514 (2570) &
  501 (2510) \\
d$_1$ \cite{ShresthaYu} &
 0.24 & 200 & 1300 & 1650 & 800 & 13.5 &  307 &
  99.6 &
  621 (3100) &
  460 (2300) \\
d$_2$ \cite{ShresthaYu} &
 0.13 & 800 & 1200 & 1650 & 1400 & 13.5 & 428 &
  87.6 &
  1000 (5690) &
  741 (4210) \\
d$_3$ \cite{ShresthaYu} &
 0.13 & 800 & 1200 & 1650 & 1400 & 13.5 & 428 &
  87.6 &
  1000 (5000) &
  741 (3710) \\
d$_4$ \cite{ShresthaYu} &
 0.86 & 30 & 1200 & 1400 & 1400 & 13.5 & 224 &
  20.8 &
  256 (1280) &
  190 (948) \\
e \cite{YeRayYi} &
 0.20 & 17.5 & 435 & 685 & 400 & 2.4 & 1580 &
  1660 &
  1590 (7930) &
  2160 (10\,800) \\
f \cite{LinTsai} &
 0.216 & 49 & 420 & 640 & 800 & 6.5 & 1860 &
  506 &
  4050 (20\,300) &
  3960 (19\,800) \\
g$_1$ \cite{ChungMiller} &
 0.1 & 62 & 450 & 700 & 250 & 8 & 1490 &
  1920 &
  5560 (27\,800) &
  5030 (25\,200) \\
g$_2$ \cite{ChungMiller} &
 0.9 & 3 & 450 & 700 & 250 & 8 & 1490 &
  154 &
  447 (2240) &
  405 (2020) \\
g$_3$ \cite{ChungMiller} &
 0.99 & 0.9 & 450 & 700 & 500 & 8 & 1490 &
  109 &
  634 (3170) &
  574 (2870) \\
h$_1$ \cite{MohammadMenon} &
 0.05 & 1000 & 450 & 750 & 2400 & 2.9 & 1670 &
  479 &
  3520 (17\,600) &
  4510 (22\,600) \\
h$_2$ \cite{MohammadMenon} &
 0.18 & 1000 & 450 & 750 & 2600 & 2.9 & 1670 &
  36.1 &
  288 (1440) &
  369 (1840) \\
h$_3$ \cite{MohammadMenon} &
 0.9 & 3.5 & 450 & 750 & 1250 & 2.9 & 1670 &
  132 &
  507 (2530) &
  650 (3250) \\
i$_1$ \cite{BanerjiMenon} &
 0.2 & 63 & 470 & 670 & 2000 & 2.9 & 1200 &
  462 &
  2880 (14\,400) &
  3690 (18\,400) \\
i$_2$ \cite{BanerjiMenon} &
 0.36 & 155 & 3000 & 5000 & 10\,000 & 2.7 & 251 &
  53.8 &
  221 (1110) &
  290 (1450) \\
i$_3$ \cite{BanerjiMenon} &
 0.81 & 2 & 560 & 800 & 1600 & 2.9 & 1010 &
  425 &
  1780 (8890) &
  2280 (11\,400) \\
j \cite{ZhangCapasso} &
 0.35 & 158 & 3000 & 4000 & 2000 & 15 & 157 &
  56.2 &
  416 (2080) &
  296 (1480) \\
k$_1$ \cite{WangTsai} &
 0.106 & 235 & 400 & 660 & 800 & 6.5 & 1860 &
  450 &
  3610 (18\,000) &
  3520 (17\,600) \\
k$_2$ \cite{WangTsai} &
 0.125 & 165 & 400 & 660 & 800 & 6.5 & 1860 &
  460 &
  3680 (18\,400) &
  3590 (18\,000) \\
k$_3$ \cite{WangTsai} &
 0.15 & 118 & 400 & 660 & 800 & 6.5 & 1860 &
  444 &
  3560 (17\,800) &
  3470 (17\,400) \\
l$_1$ \cite{BalliHastings} &
 0.278 & 34.5 & 1000 & 1800 & 5900 & 2.4 & 837 &
  424 &
  2380 (11\,900) &
  3260 (16\,300) \\
l$_2$ \cite{BalliHastings} &
 0.11 & 181 & 1000 & 1800 & 5900 & 2.4 & 837 &
  543 &
  3060 (15\,300) &
  4180 (20\,900) \\
l$_3$ \cite{BalliHastings} &
 0.06 & 665 & 1000 & 1800 & 5900 & 2.4 & 837 &
  499 &
  2810 (14\,100) &
  3840 (19\,200) \\
m \cite{YuZhang} &
 0.3 & 60 & 1000 & 1200 & 1000 & 13.5 & 314 &
  207 &
  2150 (10\,800) &
  1590 (7970) \\
n \cite{ChengZhuang} &
 0.385 & 12\,000 & 0.375 mm & 1 mm & 550 $\mu$m & 12 & 3.14 &
  0.598 &
  5.96 (29.8) &
  4.64 (23.2) \\
 \hline
\end{tabular}
}

 \caption{
  Summary of design parameters and performance values of broadband achromatic metalenses in the literature.
  Listed are the numerical aperture (NA), the focal length ($F$), the wavelength range ($\lambda_\text{min}, \lambda_\text{max}$), the device thickness ($L$), the maximum relative permittivity used in the device ($\varepsilon_{r\text{max}}$), the nominal bandwidth ($\Delta \omega$), and the three limits derived from the single-resonance, Miller's, and Tucker's time-bandwidth products, given by \eqref{eq:limitSR}, \eqref{eq:limitMiller} and \eqref{eq:modifiedTucker}, respectively.
  In parentheses in the last two columns are the relaxed bandwidth limits accounting for an error in the implemented time delay $\Delta T_\text{err} = 0.8 \Delta T$, which corresponds to large aberrations and lower focal spot intensity, as discussed in the text.
  All design parameters are quoted from the respective sources with the exception of some values of permittivity, which are estimated based on the material if not reported \cite{TiO2_1, TiO2_2, Palik, GaSb}.
 }
 \label{fig:LimitTable}
\end{table*}

\clearpage

\end{document}